\documentclass[%
 reprint,
 amsmath,amssymb,
 aps,
]{revtex4-2}

\usepackage{graphicx}% Include figure files
\usepackage{dcolumn}% Align table columns on decimal point
\usepackage{bm}% bold math
\def\eps{\epsilon}%
\def\tensor{\,\raise2pt\hbox{${}_{\otimes}$}\,}% Tensor
\def\fdg{\,:\,}% ... fuer die gilt ...
\def\ptl{\partial}% Partial
\def\rest#1{\raise-2pt\hbox{${\lfloor_{#1}}$}}% Restringiert
\def\olin#1{\overline{#1}{}}% Oben-quer
\def\ulin#1{\underline{#1}{}}% Unterstreichen
\def\halb{\frac{1}{2}}% 1/2
\newcommand{\ba}{\begin{array}}
\newcommand{\ea}{\end{array}}
\newcommand{\bea}{\begin{eqnarray}}
\newcommand{\eea}{\end{eqnarray}}
\newcommand{\bee}{\begin{eqnarray*}}
\newcommand{\eee}{\end{eqnarray*}}

\newcommand{\la}{\lambda}

\usepackage{color}

\newcommand{\green}[1]{{\color{green}#1}}

\newcounter{mnotecount}[section]

\renewcommand{\themnotecount}{\thesection.\arabic{mnotecount}}

\newcounter{mymnotecount}[section]
\renewcommand{\themymnotecount}{\thesection.\arabic{mymnotecount}}
\newcommand{\mymnote}[1]{\protect{\stepcounter{mymnotecount}}${\raisebox{0.5\baselineskip}[0pt]{\makebox[0pt][c]{\color{green}{\tiny\em$\bullet$\themnotecount}}}}$\marginpar{\raggedright\tiny\em$\!\bullet$\themymnotecount:

\green{#1}}\ignorespaces}

\renewcommand{\mymnote}[1]{}
\begin{document}

\preprint{APS/123-QED}

\title{Energy Extraction, or Lack Thereof}% Force line breaks with \\
%\thanks{A footnote to the article title}%

\author{Nishanth Gudapati}
\email{Ngudapati@clarku.edu}
\affiliation{Clark University, Department of Mathematics, 950 Main Street, Worcester, MA 01610, USA}

\date{\today}% It is always \today, today,
             %  but any date may be explicitly specified

\begin{abstract}

The problem of stability of rotating black holes is the subject of a long standing research program since the 1960s and remains an unresolved problem in general relativity.
A major obstacle in the black hole stability problem is that the energy of waves propagating through rotating black holes spacetimes is not necessarily positive-definite, due to the so called ergo-region. This is a serious complication that limits the efficacy of most mathematical techniques.  In this Letter, we report that, despite the ergo-region, there exists a positive-definite \emph{total energy} for axisymmetric Maxwell, gravitational and electrovacuum perturbations of Kerr and Kerr-Newman black hole spacetimes. 
	
	%In this Letter, we report that there exists a positive-definite energy functional for axially symmetric electromagnetic, gravitational and electrovacuum perturbations for rotating black hole spacetimes.   
\end{abstract}

%\keywords{Suggested keywords}%Use showkeys class option if keyword
                              %display desired
\maketitle

\section{Introduction}
	Einstein’s general theory of relativity was a great revelation of 20th century science that provided a
	rigorous geometric basis for ideas such as the equivalence principle and the mass. It also predicted black holes and gravitational waves. Mathematically,  the Einstein equations for general relativity can be represented as 
	\begin{align} \label{EE}
	\bar{R}_{\mu \nu} =0, \quad (\bar{M}, \bar{g})
	\end{align}
	where $\bar{R}_{\mu \nu}$ is the Ricci curvature of the $3+1$ Lorentzian manifold $(\bar{M}, \bar{g}).$
	Despite their simple appearance, in practice, these equations are a system of highly complex nonlinear partial differential equations. As a consequence, several mathematical and theoretical problems, such as the black hole stability problem, are wide open. This problem is the subject of this Letter.  
	%\subsection*{The Black Hole Stability Problem}
	
	Black holes are exotic objects in the universe that provide a scientific basis for exotic notions like wormholes, time travel, the big-bang etc.
	In a major recent development, black holes have
	been experimentally observed. Mathematically, black holes are solutions of the Einstein equations \eqref{EE} for general relativity. In order to establish the physical relevance of black holes that occur in Einstein's theory, it is crucial to establish that those black holes are stable under the perturbations of the Einstein equations (`the black hole stability problem'). In particular, the Kerr family of black holes is an important two parameter $(a, m)$  (where $a$ represents the specific angular-momentum and $m$ represents the mass)   family of solutions of the Einstein equations that forms the genesis of the black hole stability problem. 
	
	%Indeed, ever since K. Schwarzschild famously discovered 
	
	Indeed, ever since Einstein famously published his equations for general relativity \eqref{EE} in 1915 and the discovery of the Schwarzschild metric shortly thereafter, there was a general expectation that there should be a solution that represents massive, rotating black hole spacetimes. As such, attempts were made by H. Weyl, J. Ernst and A. Papapetrou to find such a solution. In 1963, when R. Kerr arrived at such a solution there was soon a wide consensus that one had arrived at the right solution. Subsequently, mathematicians and physicists conjectured that the Kerr family represents the unique asymptotically-flat solution of the stationary Einstein equations (`the black hole uniqueness problem') and that it is stable under perturbations with respect to the dynamical Einstein equations for general relativity (`the black hole stability problem'). Subsequently,
	several stalwarts in the past have contributed to the study of these problems. 
	These results are streamlined and summarized in the classical monograph of Chandrasekhar \cite{Chandrasekhar_83}. In recent times, the black hole stability problem is under intense investigation in the mathematics community (see e.g., \cite{ABBM_19, HHV_19, KS_21, HDRT_21})
	
	An important first step in the resolution of the black hole stability problem is to study the Maxwell equations and the linear perturbative theory of the Einstein equations around the Kerr black hole spacetimes. However, a peculiar property of Kerr black holes $( a \neq 0 )$ is that one is always surrounded by a so called ergo-region. In the ergo-region, an object cannot remain stationary and is forced to move along with the rotating black hole. A surprising feature of the ergo-region can be explained using Roger Penrose's gedanken experiment (see FIG. 1.). Suppose we throw an object into a Kerr black hole in such a way that it splits into two pieces (as shown in FIG. 1.), where one piece enters the black hole and the other exits the ergo-region. The piece that exits can have higher energy than the original object. This is not a violation of the conservation of energy because the piece that goes into the black hole carries `negative' energy. 
	
	This phenomenon can also be seen for linear scalar waves propagating on Kerr black holes \cite{AAS_73,FKSY_08} (see also \cite{DC_70}).  In other words,  linear scalar waves propagating on a Kerr black hole, do not necessarily have a positive-definite energy. By analogy with Penrose's gedanken experiment, this results in a phenomenon where it appears as though the black hole is emitting energy. This phenomenon is referred to as the Penrose process or superradiance or energy extraction.

	\begin{figure}
		\centering
		\caption{Penrose process or energy extraction for a rotating Kerr black hole}
		\includegraphics[scale=0.50]{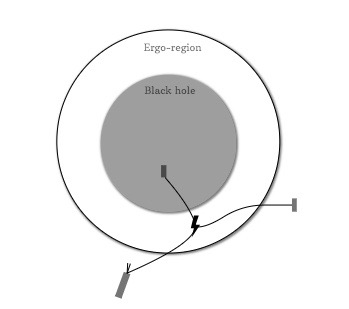}
	\end{figure}
	
	This feature of the ergo-region has important implications for the mathematical problem of the stability of Kerr black hole spacetimes. The aforementioned model of  waves propagating on a Kerr black hole is analogous to the Maxwell and  linear perturbations of the Einstein equations around a Kerr black hole spacetime. As it can be intuitively imagined, the stability of a system is closely linked to the decay of perturbations, which in turn is associated to `loss'. The fact that a Kerr black hole effectively `emits' energy, counteracts the `decaying' nature of the perturbations. In fact, in axial symmetry, it can be established that, if the energy is not positive, it will blow up exponentially in time, implying instability of Kerr black hole spacetimes \cite{WP_13}.  Furthermore, the methods of partial differential equations that establish decay are typically based on a positive energy. Likewise, in the Lyapunov theory of stability of dynamical systems, a notion of energy
	 and its positive-definiteness serve as  important criteria for various notions of stability.
	Thus, the lack of a positive-definite and conserved energy is a significant issue that limits the strength of various mathematical methods in the resolution of the black hole stability problem.

	In this Letter, we report that, despite the ergo-region, there exists a positive-definite \emph{total energy} for axisymmetric perturbations of Kerr and Kerr-Newman black hole spacetimes. 
	
%	The problem of positivity of energy can also be interpreted from an alternative perspective using the so called ADM mass. The notion of mass-energy for the Einstein equations is provided by the  ADM mass 
	%of asymptotially-flat Riemannian hypersurface $(\olin{\Sigma}, \bar{q})$:
	
%	\begin{align}
%	m_{\text{ADM}} \fdg = \frac{1}{16\pi} \lim_{r \to \infty} \int_{\mathbb{S}^2} \sum^3_{i, j=1} (\ptl_i \bar{q}_{ij} - \ptl_j \bar{q}_{ii}) \ \frac{ x^j }{r} \bar{\mu}_{\mathbb{S}^2}.
%	\end{align} 
%	It was established in the celebrated works of Schoen-Yau \cite{schoen-yau-1,schoen-yau-2} and Witten \cite{witten-pmt} that $m_{\text{ADM}} \geq 0$
%	(`the positive mass theorem'). However, it is \textbf{not necessary} that the positive mass theorem extends to the perturbative theory of Einstein's equations.
	\section{Electromagnetism in Kerr black hole spacetimes}
%\section{A Positive-Definite Energy}

	In the case of axisymmetric linear scalar waves propagating in a Kerr black hole spacetime, the problem of the ergo-region disappears on account of the fact that Kerr black holes are also axially symmetric. However, the problem reappears for both axisymmetric Maxwell and linearized Einstein fields propagating on Kerr black holes.  In fact, one can arrange for the axisymmetric Maxwell equations that their energy density is \emph{locally negative}. 
	
	The question of whether the axisymmetric Maxwell and Einstein equations admit a positive-definite total energy (for $\vert a \vert<M$), somewhat analogous to axisymmetric linear waves, has been one of those perplexing `borderline' open problems for decades.
	
	In a recent work, we were able to resolve this long standing open problem. Specifically, we were able to establish that one can construct a positive-definite \emph{total} energy for axisymmetric Maxwell fields propagating through sub-extremal Kerr black hole spacetimes ($\vert a \vert <M$)  \cite{GM17_gentitle}. An interesting aspect of our result
	is that we are able to construct a positive-definite \emph{total integrated} energy, even though the \emph{local} energy density could be \emph{negative}. Our proof relies on a Hamiltonian framework of the dimensionally reduced axisymmetric Einstein equations and certain transformations that were originally developed for the black hole uniqueness theorems by B. Carter and D.C. Robinson \cite{Car_71, Rob_74}. 
	%In view of the aforementioned complications with the ergo-region and superradiance, this result considerably simplifies the analysis of the black hole stability problem. 
	
	In the following, we shall present our proof for the construction of a positive-definite and conserved energy for axisymmetric electromagnetic (Maxwell) fields propagating on Kerr black hole spacetimes. 
	 
	 Consider the Lagrangian variational principle of the Maxwell equations: 
	\begin{align}
	S_{\text{Max}} \fdg = - \frac{1}{4} \int  \Vert F \Vert_{\bar{g}} ^2\, \bar{\mu}_{\bar{g}}, \quad 
	\text{ on the Kerr metric $(\bar{M}, \bar{g}),$ }
	\end{align}
	where $F$ is the Maxwell tensor derivable from the vector potential $A$ and $\bar{\mu}_{\bar{g}}$ is the volume form of $(\bar{M}, \bar{g})$. It is well known that \cite{MTW}, if we perform a Legendre transform of the functional $S_{\text{Max}},$ we get a Hamiltonian energy functional $H^{\text{Max}}$ over the phase space $X^{\text{Max}} \fdg = \{ A_i, \mathcal{E}^i \}$ together with the Maxwell constraints  $\ptl_i \mathcal{E}^i =0$ and $\ptl_i \mathcal{B}^i =0$, where $\mathcal{E}$ and $\mathcal{B}$ are the `electric' and `magnetic' components of $F$ respectively: 
\begin{align}\label{H-max}
H^{\text{Max}} \fdg= \int_{\olin{\Sigma}} 
\left( \halb \bar{N} \bar{q}_{ij} \bar{\mu}^{-1}_{\bar{q}} (\mathcal{E}^i \mathcal{E}^j + \mathcal{B}^i \mathcal{B}^j) - N^i \eps_{ijk} \mathcal{E}^j \mathcal{B}^k \right) d^3x
\end{align}	
	where $\bar{N}$ and $\bar{N}^i$ are the lapse and the shift, corresponding to an ADM decomposition of $(\bar{M}, \bar{g}) = (\olin{\Sigma}, \bar{q}) \times \mathbb{R}.$ It may be seen readily in \eqref{H-max} that it has an indefinite sign, even in axial symmetry. We now introduce the variables $\lambda$ and $\eta$ (`the twist potentials') such that $\mathcal{E}^a = \eps^{ab} \ptl_b \eta$ and $\mathcal{B}^a = \eps^{ab} \ptl_b \lambda,$ taking advantage of the constraint equations on the quotient space $\Sigma \fdg = \olin{\Sigma}/SO(2)$ and form the new phase space $X :$
		\begin{align}
	X \fdg = \big\{ ( \lambda, v), (\eta, u) \big\}, \quad \text{where $v \fdg = - \mathcal{E}^\phi $ and $u \fdg= \mathcal{B}^\phi.$}
	\end{align}
	 The energy functional $H^{\text{Max}}$ can now be reduced to 
	 \begin{widetext}
	\begin{align} \label{H-alt}
	H^{\text{Alt}} \fdg =& \int_{\Sigma} e^{\text{Alt}} d^2x, \quad  \text{where $e^{\text{Alt}}$ is defined as,} \\
	e^{\text{Alt}} \fdg= & \halb N e^{2 \gamma} \bar{\mu}^{-1}_q (v^2 + u^2) + \halb N \bar{\mu}_q q^{ab} e^{-2\gamma} (\ptl_a \eta \ptl_b \eta + \ptl_a \lambda \ptl_b \lambda) - \bar{N}^\phi \eps^{ab} \ptl_a \eta \ptl_b \lambda.
	\end{align}
	\end{widetext}
	Again notice that $H^{\text{Alt}}$ has an indefinite sign. In fact, we can arrange that the energy density $e^{\text{Alt}}$  is locally negative inside the ergo-region. To see this, choose $u=v=0$ and $\lambda, \eta$ such that they satisfy Cauchy-Riemann equations $\ptl_\rho \lambda = \ptl_z \eta$, $\ptl_z \lambda=- \ptl_\rho \eta $ in cylindrical coordinates for $\Sigma$. It can then be readily seen that $e^{\text{Alt}}$ is \emph{negative} inside the ergo-region. 
	
	 While the positivity problem is not immediately solved in this new phase space $X$, the advantage of this formulation is that it allows a series of transformations that result in a positive-definite, regularized, Hamiltonian (Section 2 in \cite{GM17_gentitle}), $H^{\text{Reg}} = \int_{\Sigma} e^{\text{Reg}} d^2x$, where $e^{\text{Reg}}$ is defined as:
	 \begin{widetext}
	 \begin{align} \label{postive-H-reg}
	 e^{\text{Reg}} \fdg=& \halb N \bar{\mu}^{-1}_q (\ulin{u}^2 + \ulin{v}^2) + \frac{1}{4} N \bar{\mu}_q q^{ab} \left( 2 \ptl_a \gamma \ptl_b \gamma + \halb e^{-4\gamma} \ptl_a \omega \ptl_b \omega   \right) ( \ulin{\lambda}^2 + \ulin{\eta}^2) \notag\\
	 & \halb N \bar{\mu}_q q^{ab} \Big( (\ptl_a \ulin{\la} - \halb \ulin{\eta} e^{-2\gamma} \ptl_a \omega)(\ptl_b \ulin{\la} - \halb \ulin{\eta} e^{-2\gamma} \ptl_b \omega) 
	 + (\ptl_a \ulin{\eta} + \halb  \ulin{\la} e^{-2\gamma} \ptl_a \omega)(\ptl_b \ulin{\eta} + \halb  \ulin{\la} e^{-2\gamma} \ptl_b \omega)  \Big)  
	 \end{align}
	 \end{widetext}
	 where the canonical pairs in the phase space $$ \ulin{X}\fdg=\big \{ (\ulin{\lambda}, \ulin{v} ), ( \ulin{\eta}, \ulin{u}) \big\}$$ are defined as $$\ulin{\eta} \fdg = e^{-\gamma}, \quad \ulin{u} \fdg= e^\gamma u, \quad \ulin{\la} \fdg= e^{-\gamma} \la, \quad \ulin{v} \fdg = e^\gamma v.$$
These transformations are motivated by the Carter-Robinson identities \cite{Car_71, Rob_74}. 
	 Our construction can be schematically represented as follows: 
	\begin{align} \label{canonical-transformations}
	(X^{\text{Max}}, H^{\text{Max}}) \to (X, H^{\text{Alt}}) \to (\ulin{X}, H^{\text{Reg}})
	\end{align} 
	where $H^{\text{Reg}} \geq 0.$ We would like to point out that these transformations are canonical transformations. In particular, $H^{\text{Reg}}$ is also a Hamiltonian, in addition to being positive. Actually, we can construct a divergence-free vector field density $J$ such that $J^t = e^{\text{Reg}}$ i.e., 
	
	\begin{align}
	\ptl_t J^t + \ptl_a J^a =0, \quad a= 1, 2. 
	\end{align}
	Furthermore, crucially, the flux of $J$ through the event-horizon $\mathcal{H}$ (which is a null hypersurface) also has a positive-definite sign i.e., 
	
	\begin{align}
	\text{Flux} (J, \mathcal{H}) \geq 0
	\end{align}
	which implies that the integrated energy flux of axisymmetric Maxwell fields is \emph{ingoing}, into the event-horizon. 
	
	Following our results, a positive energy functional for Maxwell fields was also constructed by Prabhu-Wald \cite{PW_17} using their notion of `canonical energy'. It turns out that the energy \eqref{postive-H-reg}  can also be used to construct a positive energy for Newman-Penrose-Maxwell scalars on Kerr black holes \cite{NG_19_1}. Furthermore, the methods presented above are versatile and they can be extended to Maxwell fields on rapidly rotating $ (\vert a \vert <M)$ Kerr-de Sitter black hole spacetimes \cite{NG_17_2}. We would like to point out that, a positive-definite energy was first constructed by Dain-de Austria \cite{DA_14} for axisymmetric linearized Einstein perturbations of extremal Kerr black holes $(\vert a \vert =M)$, using the Brill mass formula \cite{B59, D09}. The positive energy presented here is a special case of our elaborate work on axisymmetric linearized Einstein-Maxwell perturbations of Kerr-Newman black hole spacetimes \cite{GM17_gentitle} \footnote{it may be recalled that the first-order linearization of Einstein-Maxwell fields about the Kerr black hole background yields decoupled, purely Maxwellian fields on the Kerr background}.   Results analogous to the ones presented here hold for linearized Einstein perturbations of  Kerr and Kerr-Newman black hole spacetimes \cite{NG_19_2, NG_21_1, GM17_gentitle}, but the mathematical methods involved are much more delicate (see Section III). For axisymmetric linearized Einstein perturbations, we use the linearization stability methods developed by Fischer-Marsden-Moncrief \cite{FM_75, FMM_80, Mon_75}. In particular, the construction of the energy is based on the kernel of the adjoint of the dimensionally reduced constraints. These methods, together with the Carter-Robinson identities, provide an explanation for why the transformations \eqref{canonical-transformations} almost magically provide a positive-definite energy, `out of the blue'. 

	\section{Linearized Einstein and Electrovacuum Perturbations}
	%Let us now discuss to gravitational and electrovacuum perturbations of Kerr and Kerr-Newman black holes. 
	There are few fundamental differences between the gravitational perturbations and the electromagnetism on Kerr black holes. In the gravitational problem, the difficulty in the construction of a positive energy is not only due to the ergo-region - there are geometric and gauge related obstacles in addition. Consider for example the case of Schwarzschild black hole spacetimes, which do not contain the ergo-region or exhibit energy extraction phenomenon. Nevertheless, the construction of a positive-definite energy for gravitational perturbations of Schwarzschild black hole spacetimes is not trivial by any means (see e.g., \cite{Moncrief_74, GH_16, PW_17_2}).

	In spite of the issues mentioned above, we report that there exists a positive-definite, gauge-invariant and strictly conserved total energy functional for linearized Einstein and electrovacuum perturbations of Kerr and Kerr-Newman black holes for the full sub-extremal range $\vert a, Q\vert <M$.  This outcome is made possible by the special structure offered by wave maps in axial symmetry. 
	
	The Einstein equations for general relativity in axial symmetry, when represented in the Weyl-Papapetrou gauge:
	
	\begin{align} \label{wp-gauge}
	    \bar{g} = e^{-2 \gamma} g + e^{2 \gamma} (d \phi + A_\nu dx^\nu)^2, \nu = 0, 1, 2,
	\end{align}
	
	can be reduced to the $2+1$ dimensional Einstein wave map system, where the wave map $$U \fdg (M, g) \to (\mathbb{H}^2, h)$$ has the negatively curved target, the hyperbolic 2-plane $\mathbb{H}^2$. Likewise, the Einstein-Maxwell equations can be reduced to the $2+1$ Einstein-wave map system, with the complex hyperbolic plane $\mathbb{H}_{\mathbb{C}}^2$ as the target manifold for the wave map.   
	We construct an energy using the quantity
	\begin{align} \label{second-var}
	    H^{\text{Alt}} \fdg = \int_{\Sigma} (N\, D^2 \cdot H ) \, d^2 x
	\end{align}
	where $H$ is the dimensionally reduced Hamiltonian constraint on $\Sigma,$ $D$ is the variational derivative and $N$ is the dimensionally reduced lapse in the orbit space $M \fdg = \bar{M}/SO(2),$ $g$ is the metric of $M$ as in \eqref{wp-gauge}, $d^2x$ is the coordinate volume form in $\Sigma$.  Importantly, that the quantity \eqref{second-var} can serve as a candidate for the energy follows from the fact that $(N, 0)$ is the kernel of the adjoint of the dimensionally reduced constraints. This insight is derived from the theory of Linearization Stability, developed by Fischer-Marsden-Moncrief \cite{FM_75, FMM_80, Mon_75}. The quantity $H^{\text{Alt}}$ in \eqref{second-var} can further be transformed using the generalized Carter-Robinson identities (analogous to \eqref{canonical-transformations}) into a positive-definite energy functional $H^{\text{Reg}}$ \cite{GM17_gentitle, NG_19_2}. It is worth pointing out that, in the linearized Einstein problem, it becomes transparent that the underlying cause for the validity of the  Carter-Robinson identities is the negative curvature of the target manifold of the wave map, the hyperbolic 2-plane $\mathbb{H}^2$ \cite{NG_19_2}.  
	
	A positive-definite energy defined on a space-like Cauchy hypersurface is of limited significance for the stability problem if it is not strictly conserved in time or if the constituting fields are not regular functions. These two issues are coupled in the sense that the obstacles for strict conservation of energy are caused by the boundary terms that occur at the boundaries of the orbit space - the axes, the spatial infinity and the intersection of axes and the horizon; and it is precisely at these boundaries that the regularity issues arise. Furthermore, in contrast with the electromagnetic problem (Section II), the regularity and the strict conservation of energy in the linearized Einstein (i.e., gravitational) and the electrovacuum problems are further complicated by gauge related issues.  
	
	In order to overcome these issues, we formulate the initial value problem in the harmonic gauge (Lorenz-harmonic gauge for electrovacuum perturbations of Kerr-Newman black holes), where we can establish causality and regularity of the future development using standard methods, and transform the perturbed fields into the Weyl-Papapetrou gauge. Firstly, we establish that the gauge-transformation as well as the gauge-transformed fields in the Weyl-Papapetrou gauge are dynamically regular at all the boundaries. Subsequently,  we prove strict conservation of the positive-definite energy $H^{\text{Reg}}$ by establishing that all the boundary flux terms vanish dynamically in time \cite{NG_21_1, GM17_gentitle}. In this context, we take advantage of the fact that the transverse-traceless tensors vanish for our geometry and thus the elliptic operators can be reduced to Poisson operators, which in turn provide the needed regularity and decay rates at the axes and infinity \cite{GM17_gentitle, NG_21_1}.

	\section{Summary and Concluding Remarks}
	The problem of stability of Kerr black holes is a major open problem in theoretical and mathematical studies in general relativity. Arguably, the most important obstacle in establishing the stability of Kerr black holes  is the ergo-region and its effects like superradiance and energy-extraction. The effects of the ergo-region are even stronger for rapidly rotating sub-extremal Kerr black holes $(\vert a \vert<M)$. Due to this reason, most of the current and past works on the black hole stability problem are dedicated to the study of stability of Kerr black holes for small angular-momentum $(\vert a \vert \ll M)$. 
	
	Theoretically, due to the energy extraction phenomenon, the energy of perturbations can blow up in time, implying instability of Kerr black holes. In this Letter, we presented a result wherein we constructed a positive-definite energy for axisymmetric Maxwell fields propagating through Kerr black holes. Analogous results hold for axisymmetric gravitational and electrovacuum perturbations of Kerr and Kerr-Newman black hole spacetimes respectively, using much more delicate mathematical methods. These results imply a form of stability of Kerr and Kerr-Newman black hole spacetimes in axial symmetry.  Since our results hold for the entire sub-extremal range of Kerr and Kerr-Newman black holes  \footnote{We would like to point out that rapidly rotating, but sub-extremal, black holes are physically important}, these results are expected to be a major step forward towards the resolution of the black hole stability problem. Furthermore, the methods are suitable for the study of stability of Kerr-Newman-de Sitter black holes and higher-dimensional black hole spacetimes under a suitable symmetry class. 
%\section{Acknowledgements}
%The author gratefully acknowledges the conversations with A. Ashtekar in the initial stages of this work. 

% The \nocite command causes all entries in a bibliography to be printed out
% whether or not they are actually referenced in the text. This is appropriate
% for the sample file to show the different styles of references, but authors
% most likely will not want to use it.
%\nocite{*}

\bibliography{central-bib(Harvard)}% Produces the bibliography via BibTeX.

\end{document}